\documentclass[a4paper,twocolumn,showpacs]{revtex4-1}

\usepackage{amsmath,amssymb}
\usepackage{amsthm}
\usepackage{latexsym}
\usepackage[dvipdf]{graphicx}
\usepackage{mathrsfs}

\newcommand{\ket}[1]{\left| #1 \right\rangle} 
\newcommand{\bra}[1]{\left\langle #1 \right|} 

\newcommand{\ketbra}[2]{\ket{#1}\!\bra{#2}}

\newcommand{\lie}{{\cal L}}

\newcommand{\tomega}{{\tilde{\omega}}}
\newcommand{\hc}{{\mbox{h.c}}}
\newcommand{\coea}[2]{a^{(#1)}_{#2}}

\begin{document}

\title{\bf A control protocol of finite dimensional quantum systems}

\author{Jianju Tang}
\author{H. C. Fu}
\email{E-mail: hcfu@szu.edu.cn, corresponding author.}
\affiliation{School of Physical Sciences and Technology, Shenzhen University,
Shenzhen 518060, P. R. China}

\begin{abstract}
An exact and analytic control protocol of two types of finite dimensional quantum
systems is proposed. The system can be drive to an arbitrary target state
using cosine classical fields in finite cycles. The control parameters which are
time periods of interaction between systems and control fields  in each cycles
are connected with the probability amplitudes of target states via triangular
functions and can be determined analytically.
\end{abstract}


\maketitle

\section{Introduction}

Quantum control is to drive a quantum system from an initial state to an arbitrary
target state through its interaction with classical control fields or with a quantum
accessor. It was first proposed by Huang {\it et.\,al.} in 1983 \cite{huang} and then
attracted much attention of chemists, physicists and control scientists.  Various notations
in classical control theory were generalized to the quantum control, such as open and
closed control, optimal control \cite{optimal}, controllability \cite{controllability,fu1,fu2,
graph}, feedback control \cite{feedback} and so on. Coherent and incoherent (indirect)
control schemes are proposed. In later case the system is controlled by its interaction
with a quantum accessor which is controlled by classical fields \cite{indirect1,indirect2,
romano,penchen}.
Typically, in the approach of quantum control, one should first model the controlled system and examine its controllability
which is determined by the system Hamiltonian and interaction Hamiltonian with
classical fields,
and then design classical fields to stream the system to the given target state, which is
referred to as the {\it control protocol} and is the issue we would like to address in
this paper. Some works were proposed along this line, for example, using the Cartan
decomposition of Lie groups \cite{cartan}.

In this paper we shall develop an explicit control protocol of finite quantum system
with (1) all distinct energy gaps such the Mores potential, and (2) all equal energy
gaps except the first one.
We use cosine classical field to drive the quantum system to arbitrary target states
in finite cycles and the control parameters are interaction time intervals between
system and control field in each cycles. Control parameters are linked with probability
amplitudes of the target states through triangular functions and can be obtained
analytically.

This paper is organized as follows. In Sec. II, we formulate the controlled system
and control scheme and investigated the controllability. In Sec. III, we present
the control protocol of system with all distinct energy gaps and in Sec. IV we
consider the system with equal energy gaps except one. We conclude in Sec.V.

\section{Control systems}

\subsection{Control Systems}

Consider an $N$-dimensional non-degenerate quantum system with eigen energy $E_n$ and corresponding
eigenstates $|n\rangle$, described by the Hamiltonian
\begin{equation}\label{eq2.1}
    H_0=\sum^N_{n=1} E_n \ket{n}\!\bra{n}.
\end{equation}
Without losing generality, we assume $H_0$ is traceless, namely $\mbox{tr}H_0 = 0$.
In this paper, we only consider two different types of systems, the first one
having all equal energy gaps except the first one, namely
\begin{equation}
\mu_1 \neq \mu_2=\mu_3=\cdots = \mu_{N-1},
\end{equation}
where $\mu_i = E_{i+1}-E_i$ is the energy gap, and another one with
all district energy gaps
\begin{equation}
\mu_i \neq \mu_j, \qquad i\neq j = 1,2,\cdots, N-1,
\end{equation}
For later convenience, we call them the system I and system II, respectively.
For System I, we also define energy gaps
\[
\tilde{\mu}_i = E_{i+1}-E_1, \qquad i=1,2,\cdots,N-1.
\]

The purpose of this paper is to develop a control scheme to drive the systems to an
arbitrary target state from an initial state, using some independent classical fields
$f_m(t)$. The total Hamiltonian of the system and control fields can be
generally written as
\begin{align}
H=H_0+H_I,
\qquad
H_I =  \sum_{m=1}^M f_m(t)\hat{H}_m,
\end{align}
where $M$ is the number of  independent classical fields.

\begin{figure}[t]
  \includegraphics[width=8cm]{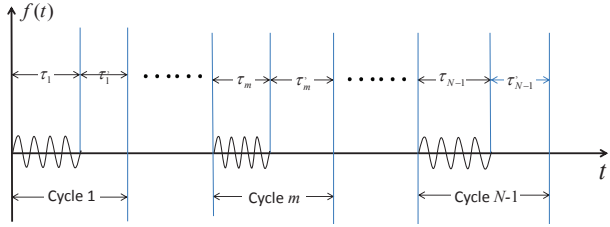}
  \caption{Control fields.}  \label{fig1-field}
\end{figure}

For the $N$-dimensional systems considered in this paper, the total control
process includes $N-1$ cycles.
In the $m$-th cycle, we first apply a classical field
\[
f_m(t)=\xi_m \cos(\nu_{m} t),
\]
where $\nu_m$ is the frequency, to control the system for time period $\tau_m$,
and then turn off the control field such that the system evaluates freely for a time
period $\tau_m^\prime$, as
showed in Fig.\ref{fig1-field}.

For system I, the frequency of the control field is chosen as
\begin{equation}
\nu_m = \tilde{\omega}_m,
\qquad
\tilde{\omega}_m \equiv \left(E_{m+1}-E_1\right)\hbar^{-1}.
\end{equation}
It is easy to find that $\tomega_m \neq \tomega_n$ for $m\neq n$. This means in
each cycle, only transition between $E_m$ and $E_1$ occurs.

For system II, the frequency $\nu_m$
is chosen as
\begin{equation}
\nu_m = \omega_m, \qquad \omega_m = \left(E_{m+1}-E_m\right)\hbar^{-1},
\end{equation}
This means that, in each cycle, only transition between level $E_m$ and $E_{m+1}$
occurs, as  all $\omega_m$ are different. So the control process includes $N-1$
cycles.

For both systems, there are two processes in each cycle. We first apply the control
field to control the system for a time period $\tau_m$, and then turn off the control
field and allow the system to evaluate for a time period $\tau_m^\prime$. We will
see that the first process provides the real probability amplitude of the target
state and the second process provides the phases. Therefore, the control field can
be rewritten as
\begin{equation}
f_m(t)=\left\{\begin{array}{ll}
\xi_m \cos(\nu_mt), & t_{m-1} \leq t \leq  t_{m-1}+\tau_m, \\
0, & \mathrm{otherwise},
\end{array}
\right.
\end{equation}
where
$t_m = \sum_{k=1}^{m}(\tau_k +\tau_k^\prime)$.
Those $N-1$ control fields are independent in the sense that each $f_m(t)\neq 0$ in
different time period.

The whole control process can be equivalently regarded as control by one
control field $f(t)=\sum_{m=1}^{N-1} f_m(t)$, where $f(t)$ is shown in
Fig.1.

\begin{figure}[t]
  \includegraphics[width=8cm]{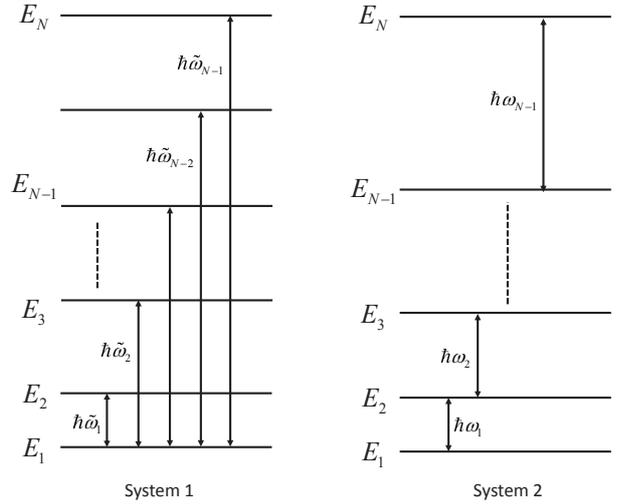}
  \caption{Interaction between systems and control fields in each cycle. For system I,
  control field causes transition between the level $E_1$ and level $E_{m+1}$,
  while for system II, the control field causes transition between adjacent energy levels.
  }
  \label{fig2}
\end{figure}

For system I, in the $m$-th cycle, we apply the classical field
\begin{equation}\label{eq3.2}
     f^{(m)}(t)=\xi_m \cos\left(\nu_{m} t\right), \qquad \nu_m = \tomega_m,
\end{equation}
which causes transition between energy levels $E_1$ and $E_{m+1}$. So
the interaction Hamiltonian between control field and system in Schr\"{o}ding picture is
\begin{align}\label{interaction-hmiltonian-system11}
  H^{(m)}=&  f^{(m)}(t) g_m \left(\ket{1}\!\bra{m+1}+ \ket{m+1}\!\bra{1}\right)
             \nonumber \\
    = & \Omega_m\cos(\nu_m t) \left(\ket{1}\!\bra{m+1}+\ketbra{m+1}{1}\right),
\end{align}
where $g_m$ is the coupling constant and $\Omega_m=\xi_m g_m$.

For system II, the interaction Hamiltonian between the system and the control
field is
\begin{align}
  H^{(m)}=& f^{(m)}(t) g_m \left(\ket{m}\bra{m+1}+\ket{m+1}\bra{m}\right)\nonumber \\
  =&\Omega_m\cos(\nu_m t)\left(\ket{m}\bra{m+1}+\hc\right),
  \label{interaction-hmiltonian-system12}
\end{align}
where $g_m$ and $\Omega_m=\xi_m g_m$ as in system I, but frequency
$\nu_m = \omega_m$.

Interaction between systems and controls are illustrated in Fig.(\ref{fig2}).

\subsection{Complete controllability}

Before presentation of the control protocol, we first examine the controllability of this
control scheme, namely, to examine whether the Lie algebra generated by
the skew-Hermitian operators $iH_0$ and $iH_m$
\begin{equation}
  \lie=\mathrm{Gen}\{ iH_0,iH_m|m=1,2,...,N-1\}
  \label{eq2.4}
\end{equation}
is $\mathrm{su}(N)$ \cite{controllability}. Here $H_m$ for system I and II
are
\begin{align}
& H_m = \ketbra{1}{m+1}+\ketbra{m+1}{1}, \\
& H_m = \ketbra{m}{m+1}+\ketbra{m+1}{m}
\end{align}
respectively.
Or equivalently, $iH_0, iH_m$ generate the Chevalley basis of su($N$) \cite{fu1,fu2}
\begin{align}
   & ix_n=i(\ket{n}\!\bra{n+1}+\ket{n+1}\!\bra{n}),\nonumber \\
   & iy_n=\ket{n}\!\bra{n+1}-\ket{n+1}\!\bra{n},\nonumber \\
   & ih_n=i(\ket{n}\!\bra{n}-\ket{n+1}\!\bra{n+1}),
\end{align}
where $n=1,2,...,N-1$. In fact, it is enough to prove $ix_n\in\lie$ (or $iy_n\in\lie$),
as $iy_n = \mu_n^{-1}[iH_0, ix_n]$ and $ih_n=-[ix_n,iy_n]/2$.

For system I, it is obvious that
\begin{align}
     ix_1&=iH_1= i(\ketbra{1}{2}+\ketbra{2}{1})\in \lie.
\end{align}
Then
\begin{align}
   iy_2&=[iH_2,iH_1]
        =\ketbra{2}{3}-\ketbra{3}{2}\in\lie, \nonumber  \\
  ix_2&=\frac{1}{\mu_2}[[iH_2,iH_1],iH_0]=i(\ketbra{2}{3}+\ketbra{3}{2})\in\lie. 
\end{align}
Recursively, we have
\begin{align}\label{eq2.11}
   iy_{m}&=\mu_m^{-1}\left[[iH_m,iH_{m-1}],iH_0\right].
\end{align}
So the system I is completely controllable.

For System II, $iH_m$ itself
\begin{align}
iH_m = i(\ketbra{m}{m+1}+\ketbra{m+1}{m}) ,
\end{align}
is nothing but the generator $ix_m$.
Therefore, the system II is completely controllable.

\section{Control protocol of System I}

In this section we investigate the control protocol of System I.
Suppose the system is initially on the ground state $\ket{\psi_0}=\ket{1}$.
The system is driven to an arbitrary target state after $N-1$ cycles.
Relationship between the control parameters $\{\tau_m, \tau_m^\prime\}$
and probability amplitude of target states is explicitly established.

\subsection{Interaction Hamiltonian}

Changing to the interaction picture, we obtain
\begin{align}\label{eq3.7}
  H^{(m)}_I= &U_0^\dagger(t)\left[\Omega_m\cos\left(\nu_m t\right)\right. \nonumber \\
  &\left. \times\left(\ket{1}\bra{m+1}+\ket{m+1}\bra{1}\right)\right] U_0(t)\nonumber \\
=&\frac{\Omega_m}{2} \left(e^{i(\nu_m-\tilde{\omega}_m)t}\ket{1}\bra{m+1} \right. \nonumber \\
 & + e^{i(\tilde{\omega}_m+ \nu_m)t}\ket{m+1}\bra{1} \nonumber \\
 & + e^{-i(\nu_m+\tilde{\omega}_m)t}\ket{1}\bra{m+1} \nonumber \\
 & \left. +e^{i(\tilde{\omega}_m-\nu_m)t}\ket{m+1}\bra{1}\right),
\end{align}
where $U_0(t)=e^{-iH_0 t/\hbar}$, and we have used
$\cos(\nu_m t)=(e^{i\nu_m t}+e^{-i\nu_m t})/2$.
As we require that the control field is resonant with
the levels $E_1$ and $E_m$, namely, $\nu_m =\tilde{\omega}_m$,
we can neglect the  high-oscillating
terms $e^{\pm i(\tilde{\omega}_m+\nu_m) t}$ under the rotating wave approximation \cite{b}.
We finally obtain
\begin{equation}\label{eq3.9}
     H^{(m)}_I=\frac{\Omega_m}{2}\left( \ket{1}\bra{m+1}+ \ket{m+1}\bra{1}\right),
\end{equation}
which does not depend on time $t$ explicitly. So the time evolution operator
in interaction picture can be written as
\begin{equation}\label{3.14}
    U^{(m)}_I(t)=e^{-iH^{(m)}_It/\hbar}.
\end{equation}
From the fact
\begin{align}
 & \left(H^{(m)}_I\right)^{2n}=\left[\frac{\Omega_m}{2}\right]^{2n}
      \left(\ket{1}\bra{1}+\ket{m+1}\bra{m+1}\right),\nonumber \\
      & \qquad n>0, \nonumber \\
 &\left(H^{(m)}_I\right)^{2n+1}=\left[\frac{\Omega_m}{2}\right]^{2n+1} \left( \ket{1}\bra{m+1}+\hc
       \right),    \nonumber \\
 & \qquad n\ge 0,  \nonumber
\end{align}
we obtain the time evolution operator as
\begin{align}
&   U^{(m)}_I(t) = I+\left(\cos(\Omega'_mt)-1\right)\left(\ketbra{1}{1}+\ketbra{m+1}{m+1}\right)\nonumber \\
& \quad -i\sin(\Omega'_mt)\left( \ketbra{1}{m+1}+\hc\right),
   \label{evolution-operator-for-system1}
\end{align}
where $\Omega'_m=\Omega_m /2\hbar$.


\subsection{Cycle 1}

Suppose that the system is initially on the state $\ket{1}$.
When interacting with control field for time period $\tau_1$,
the system is on the state
\begin{align}\label{3.19}
   \ket{\psi_1}_I=&U_I(\tau_1)\ket{1}_I \nonumber \\
                 =&\cos(\Omega'_1 \tau_1)\ket{1}-i\sin(\Omega'_1 \tau_1)\ket{2}.
\end{align}
Changing back to the Schr\"{o}ding picture, we have
\begin{align}
   \ket{\psi_1}_S=
   & e^{-iE_1\tau_1/\hbar}\cos(\Omega'_1\tau_1)\ket{1} \nonumber \\
     & -i e^{-iE_2\tau_1/\hbar}\sin(\Omega'_1\tau_1) \ket{2}.
     \label{3.21}
\end{align}
We then turn off the external field and allow the system to evaluate for time period $\tau'_1$.
We get
\begin{align}
   \ket{\psi_1}'_S=e^{-iH_0\tau'_1/\hbar}\ket{\psi_1}_S
                 = a^{(1)}_{1}\ket{1}+a^{(1)}_{2}\ket{2}, \label{cycle1-final}
\end{align}
with
\begin{align}
  & a^{(1)}_{1}=e^{-iE_1(\tau_1+\tau'_1)/\hbar}\cos(\Omega'_1\tau_1), \nonumber \\
  & a^{(1)}_{2}=-e^{-iE_2(\tau_1+\tau'_1)\hbar}\,i\sin(\Omega'_1\tau_1) .
  \label{cycle-2-12}
\end{align}

\subsection{Cycle 2}

For the cycle 2, the initial state is the the final state of cycle 1,
namely, the state (\ref{cycle1-final}).
We first apply the control field $f_2(t)=\xi_2 \cos(\nu_2 t)$ for time
period $\tau_2$.
Using \eqref{evolution-operator-for-system1} for $m=2$, we obtain
the state in the interaction picture
\begin{align}\label{3.28}
\ket{\psi_2}_I=&U_I(\tau_2)\ket{\psi_1}'_I    \nonumber \\
=& \cos(\Omega'_2\tau_2)a^{(1)}_{1}\ket{1}+a^{(1)}_{2}\ket{2}  
  -i\sin(\Omega'_2\tau_2) a^{(1)}_{1}\ket{3}.
\end{align}
Changing back to Schr\"{o}dinger picture, and after free evolution for time period $\tau'_2$,
the state is
\begin{align}
\ket{\psi_2}'_S= & U_0(\tau_2^\prime)\ket{\psi_2}_S
=U_0(\tau_2^\prime) U_0(\tau_2)\ket{\psi_2}_I \nonumber \\
= & a^{(2)}_{1}\ket{1}+a^{(2)}_{2}\ket{2}+a^{(2)}_{3}\ket{3},
\end{align}
where
\begin{align}
     a^{(2)}_{1}&=e^{-iE_1(\tau_2+\tau'_2)/\hbar}\cos(\Omega'_2\tau_2)\,a^{(1)}_{1},    \nonumber \\
     a^{(2)}_{2}&=e^{-iE_2(\tau_2+\tau'_2)/\hbar}\,a^{(1)}_{2},    \nonumber \\
     a^{(2)}_{3}&=-e^{-iE_3(\tau_2+\tau'_2)/\hbar}\,i\sin(\Omega'_2\tau_2) a^{(1)}_{1}.
     \label{3.32}
\end{align}

\subsection{From $(m-1)$-th to $m$-th cycle}

To obtain the explicit expression of the target state, we need the recursion
relation between coefficients of the $m$-th cycle and $(m-1)$-th cycle.
Suppose that after the $(m-1)$-th cycle, we obtain the state
\begin{equation}
   \ket{\psi_{m-1}}'_S=\sum_{k=1}^m a^{(m-1)}_k \ket{k}. 
   \label{3.34}
\end{equation}
Interacting with control field for time period $\tau_m$, we have the state 
in Schr\"{o}ding picture
\begin{align}
&   \ket{\psi_m}_S  
           =e^{-iE_1\tau_m/\hbar}\cos(\Omega'_{m}\tau_m)\,a^{(m-1)}_1\ket{1}   \nonumber \\
&  \quad  + \sum_{k=2}^m a^{(m-1)}_k e^{-iE_k \tau_m/\hbar} \ket{k}   \nonumber \\
&  \quad  -e^{-iE_{m+1}\tau_m/\hbar}\,i\sin(\Omega'_{m}\tau_m) a^{(m-1)}_1 \ket{m+1}. \nonumber
\end{align}
After free evolution for time period $\tau^\prime_m$, the final state of the $m$-th cycle is
\begin{align}
\ket{\psi_m}'_s = e^{-iH_0\tau'_m/\hbar}\ket{\psi_m}_S
=\sum_{k=1}^{m+1} a^{(m)}_k \ket{k}, \label{final-state-m-cycle}
\end{align}
where the coefficients are
\begin{align}
       a^{(m)}_{1}&=e^{-iE_1 T_m /\hbar}
                \cos(\Omega'_{m}\tau_m) a^{(m-1)}_1,  \label{recursion-1}   \\
       a^{(m)}_{k}&=e^{-iE_k T_m /\hbar}\, a^{(m-1)}_k,  \quad 2\leq k \leq m,   \label{recursion-2}   \\
       a^{(m)}_{m+1}&= - e^{-iE_{m+1} T_m /\hbar}\,
       i\sin(\Omega'_{m}\tau_m)  a^{(m-1)}_1,   \label{recursion-3}
\end{align}
and we use the notation
\begin{align}
T_m \equiv \tau_m + \tau_m^\prime
\end{align}
hereafter. Eqs.(\ref{recursion-1}-\ref{recursion-3}) establishes the relationship between the probability
amplitudes of the $(m-1)$-th cycle and the $m$-th cycle.

\subsection{Target state}

For 2 and 3 dimensional system, the target state has been found in subsection B and C. So
we suppose that $N\geq 4$ in the rest of this section. From (\ref{recursion-1}) and
(\ref{cycle-2-12}), we can
easily find that
\begin{align}
a^{(m)}_{1}=\exp\left[-\frac{iE_1}{\hbar}\sum^m_{i=1}T_i\right]
         \prod^m_{i=1}\cos\left(\Omega'_{i}\tau_i\right).
         \label{coefficient-m-1}
\end{align}
As $a_{m+1}^{(m)}$ depends on $a^{(m-1)}_1$ only, we can easily find that
\begin{align}
a^{(m)}_{m+1}&=-\exp \left\{ -\frac{i}{\hbar}\left[E_{m+1} T_m +
           E_1\sum^{m-1}_{i=1} T_i \right]\right\}  \nonumber \\
         & \quad \ \times i\sin(\Omega'_{m}\tau_m)
           \prod^{m-1}_{i=1}\cos(\Omega'_{i}\tau_i).
           \label{coefficient-m-m+1}
\end{align}
For other coefficients, using \eqref{3.32}, we obtain the explicit probability amplitude
\begin{align}
a^{(m)}_{2}=&-\exp\left[-\frac{iE_2}{\hbar}\sum^m_{i=1}T_i \right]\,i
         \sin(\Omega'_{1}\tau_1), \label{coefficient-m-2} \nonumber\\ %
a^{(m)}_{3}=&-\exp\left\{-\frac{i}{\hbar}\left[E_3\sum^m_{i=2}T_i+
         E_1 T_1 \right]\right\}\nonumber\\
& \times i\sin(\Omega'_{2}\tau_2)\cos(\Omega'_{1}\tau_1).
\end{align}
To derive coefficient $a^{(m)}_{k}$, $3\leq k \leq m$, we use
\begin{align}
a^{(m-1)}_{m}= & -\exp \left\{-\frac{i}{\hbar}\left[E_m T_{m-1}+E_1\sum^{m-2}_{i=1}T_i\right] \right\} \nonumber \\
&\times i\sin(\Omega^\prime_{m-1}\tau_{m-1}) \prod^{m-2}_{i=1}\cos(\Omega^\prime_{i}\tau_i),
\label{3.42}
\end{align}
which is obtained from (\ref{coefficient-m-m+1}) by replacing $m$ by $m-1$.
Then we can recursively have
\begin{align}
a^{(m)}_{m}=& e^{-iE_m T_m /\hbar}\,a^{m-1}_{m}
       = \mathrm{exp}\left[-\frac{i E_m}{\hbar}
          \sum^m_{i=m}T_i \right]\,a^{m-1}_{m}     \nonumber\\
       =& -\exp\left\{-\frac{i}{\hbar}\left[E_m \sum^m_{i=m-1}T_i+
          E_1\sum^{m-2}_{i=1}T_i \right]\right\} \nonumber \\
        & \times i\sin(\Omega^\prime_{m-1}\tau_{m-1})\,
          \prod^{m-2}_{i=1}\cos(\Omega\prime_{i}\tau_i).
\label{3.43}
\end{align}
According to \eqref{3.42} and \eqref{3.43}, we can obtain $a^{m}_{k}$
\begin{align}
a^{m}_k
=& e^{-iE_kT_m/\hbar} a_k^{(m-1)}\nonumber \\
=& e^{-iE_kT_m/\hbar} e^{-iE_kT_{m-1}/\hbar} a_k^{(m-2)}\nonumber \\
=& \cdots\nonumber \\
=& e^{-iE_k(T_m+T_{m-1}+\cdots+T_{k+1})} a^{(k)}_k\nonumber \\
=& -\exp\left\{-\frac{i}{\hbar}\left[E_k\sum^{m}_{i=k-1}T_i
                + E_1\sum^{k-2}_{i=1}T_i\right]\right\}   \nonumber \\
          &    \times i\sin(\Omega'_{k-1}\tau_{k-1})
            \prod^{k-2}_{i=1}\cos(\Omega'_{i}\tau_i), \nonumber \\
          & (3\leq k\leq m),  \label{coffficient-m-k}
\end{align}
where we have used the result of $a^{(k-1)}_k$ given by (\ref{coefficient-m-m+1})
with $m$ replaced by $k$. Notice that when $k=m+1$, (\ref{coffficient-m-k})
recovers the $a^{(m)}_{m+1}$ given in (\ref{coefficient-m-m+1}). Therefore
(\ref{coffficient-m-k}) is also valid for $k=m+1$, and all the probability
amplitudes after $m$-th cycle are given by (\ref{coefficient-m-1}),
(\ref{coefficient-m-2}) and (\ref{coffficient-m-k}) with $3\leq k \leq m+1$.

For the system we considered here with dimension $N$, we need $N-1$ cycles to
arrive at arbitrary target states. Letting $m=N-1$, we obtain the probability
amplitude of the target state
\begin{align}
a^{(N-1)}_{1}=& \exp\left[-\frac{iE_1}{\hbar}\sum^{N-1}_{i=1}T_i\right]
            \prod^{N-1}_{i=1}\cos(\Omega'_{i}\tau_i),   \nonumber \\
a^{(N-1)}_{2}=& -\exp\left[-\frac{iE_2}{\hbar}\sum^{N-1}_{i=1}T_i\right]\,
            i\sin(\Omega'_{1}\tau_1),   \nonumber \\
a^{(N-1)}_{k}=& -\exp\left\{-\frac{i}{\hbar}\left[E_k\sum^{N-1}_{i=k-1}T_i
                + E_1\sum^{k-2}_{i=1}T_i\right]\right\}   \nonumber \\
          &    \times i\sin(\Omega'_{k-1}\tau_{k-1})
            \prod^{k-2}_{i=1}\cos(\Omega'_{i}\tau_i), \nonumber \\
          & (3\leq k\leq N).
\label{3.44}
\end{align}

\subsection{Control parameters}

For a control problem, the target state, or in other words, the
amplitude $a^{N-1}_m$ of the target state, is given. What we need to do is to
determine the control parameters
$\{ \tau_i, \tau'_i \ |\ i=1,2,...,N-1 \}$ from the probability amplitude
of the target state. For convenience, we write the target state as
\begin{equation}\label{3.45}
     \ket{\psi}=\sum^N_{n=1}\gamma_n C_n \ket{n}
\end{equation}
where $C_n$s are the real part of the amplitude
\begin{align}
     C_1&=\prod^{N-1}_{i=1}\cos(\Omega'_{i}\tau_i), \label{coefficient-c-1}\\
     C_2&=\sin(\Omega'_1\tau_1),  \label{coefficient-c-2} \\
     C_n&=\sin(\Omega'_{n-1}\tau_{n-1})\prod^{n-2}_{i=1}\cos(\Omega'_{i}\tau_i)\ ,\ 3\leq n\leq N,
      \label{coefficient-c-3}
\end{align}
and $\gamma_n$s are phases
\begin{align}
   \gamma_1 = & \exp\left[-\frac{iE_1}{\hbar}\sum^{N-1}_{i=1}T_i\right],\\
   \gamma_2 = & -i\exp\left[-\frac{iE_2}{\hbar}\sum^{N-1}_{i=1}T_i\right],\\
   \gamma_n = & -i\exp\left\{-\frac{i}{\hbar}
   \left[E_n\sum^{N-1}_{i=n-1}T_i
   +E_1\sum^{n-2}_{i=1}T_i\right]\right\},\nonumber \\
    & (3\leq n\leq N). \label{coefficient-phase}
\end{align}

For a given target state, namely, $C_n$ and $\gamma_n$ are given,
we can calculate control parameters $\{\tau_n,\tau'_n|n=1,2,...,N-1  \}$.
From (\ref{coefficient-c-2}) and $C_2$, we can determine $\tau_1$.
Then form (\ref{coefficient-c-3}) with $n=3$, we can obtain
$\tau_2$ from $C_3$. Repeating this process, we can obtain
all parameters $\tau_n$, $n=1,2,...,N-1$ from (\ref{coefficient-c-3}).

All $\tau^\prime_i$ can be obtained from (\ref{coefficient-phase}).
From $\gamma_2$ and $\gamma_3$, we can obtain
\begin{equation}
\sum_{i=1}^{N-1} T_i, \qquad E_3 \sum_{i=2}^{N-1}T_i + E_1 T_1,
\label{coefficient-003}
\end{equation}
from which we find $T_1$ and $\sum_{i=2}^{N-1}T_i$. From $\gamma_4$, we
find
\begin{equation}
E_4 \sum_{i=3}^{N-1}T_i + E_1(T_1+T_2),
\end{equation}
from which as well as  $T_1$ and $\sum_{i=2}^{N-1}T_i$, we can obtain $T_2$.
Repeating this process, we can obtain all $T_i$ and thus all $\tau^\prime_i$.

\section{Control Protocol of System II}

\subsection{Time evolution operator}

For system II, the interaction Hamiltonian is given in
Eq.(\ref{interaction-hmiltonian-system12}). This Hamiltonian
is same as (\ref{interaction-hmiltonian-system11}) for system I except the state $\ket{1}$ is
replaced by $\ket{m}$. So we can follow exactly the same
procedure as in last section, namely, changing to the
interaction picture, using rotating wave approximation, and
obtaining a time-independent Hamiltonian in interaction
picture
\begin{equation}\label{eqb3.9}
     H^{\prime (m)}_I=\frac{\Omega_m}{2}\left( \ket{m}\bra{m+1}+
                 \ket{m+1}\bra{m}\right).
\end{equation}
Using
\begin{align}
& \left(H^{\prime(m)}_I\right)^{2n}=\left(\frac{\Omega_m}{2}\right)^{2n}
  \left(\ket{m}\bra{m}+\ket{m+1}\bra{m+1}\right) \nonumber \\
& \qquad (n>0)  \nonumber \\
& \left(H^{\prime(m)}_I\right)^{2n+1}=\left(\frac{\Omega_m}{2}\right)^{2n+1}
  \left( \ket{m}\bra{m+1}\right. \nonumber \\
& \left. \qquad + \ket{m+1}\bra{m}\right),\qquad(n\ge0), \nonumber
\end{align}
we can obtain the time evolution operator in the interaction picture
\begin{align}\label{eq3.16}
&   U_I^{(m)}(t)= I+[\cos(\Omega'_mt)-1](\ket{m}\!\bra{m}+\ket{m+1}\!\bra{m+1}) \nonumber \\
&   \ \  -i\sin(\Omega'_mt)( \ket{m}\!\bra{m+1}+ \ket{m+1}\!\bra{m}),
\end{align}
where $\Omega'_m= \Omega_m/(2\hbar)$.

\subsection{Determine amplitude $a_m$}

For this model, the cycle 1 is exactly the same as the system I.
So after the cycle 1, the system is driven to the state
\begin{align}\label{finalStaWOF1}
\ket{\psi_1}'_S=a^{(1)}_{1}\ket{1}+a^{(1)}_{2}\ket{2},
\end{align}
where
\begin{align}\label{initialaCondition}
   a^{(1)}_{1}=&e^{-iE_1T_1/\hbar}\cos(\Omega'_1\tau_1),\nonumber \\
   a^{(1)}_{2}=&-e^{-iE_2T_1/\hbar}\,i\sin(\Omega'_1\tau_1).
\end{align}

Different from system I, in cycle 2, the control field $f(t)=\xi_2\cos(\omega_2t)$ causes transition
between $\ket{2}$ and $\ket{3}$. We can find the state after the cycle 2 in
Schr\"{o}dinger picture as
\begin{equation}\label{shortForm2}
  \ket{\psi_2}'_S=
  a^{(2)}_{1}\ket{1}+a^{(2)}_{2}\ket{2}+a^{(2)}_{3}\ket{3},
\end{equation}
where
\begin{align}\label{coefficient2}
  a^{(2)}_{1}=&e^{-iE_1 T_2/\hbar}a^{(1)}_{1},  \\
  a^{(2)}_{2}=&e^{-iE_2 T_2/\hbar}\cos(\Omega'_2\tau_2)a^{(1)}_{2}\\
  a^{(2)}_{3}=&-e^{-iE_3 T_2/\hbar}i\sin(\Omega'_2\tau_2) a^{(1)}_{2}.
\end{align}

To obtain the target state, we first find the recursion relations
between the $(m-1)$-th cycle and the $m$-th cycle. To this end,
we suppose that, after $m-1$ cycles, the system is on the state
\begin{equation}\label{State-ksteps}
  \ket{\psi_{m-1}}'_S= \sum_{k=1}^m a^{(m-1)}_{k}\ket{k}.
\end{equation}
Then after interactions with the control field for
time period $\tau_m$, and free evolution for time period $\tau_m^\prime$,
we find the final state after cycle $m$ as
\begin{equation}\label{shortform-k}
  \ket{\psi_m}'_S=\sum_{k=1}^{m+1} a^{(m)}_k \ket{k},
\end{equation}
with ($m \geq 2$)
\begin{align}
  &a^{(m)}_k=e^{-iE_k T_m/\hbar}a^{(m-1)}_k, \quad 1\leq k\leq m-1, \label{induction-model2-1} \\
  &a^{(m)}_{m}= e^{-iE_m T_m/\hbar}\cos(\Omega'_{m}\tau_m) a^{(m-1)}_m,  \\
  &a^{m}_{m+1}=-e^{-iE_{m+1}T_m/\hbar}i
  \sin(\Omega'_{m}\tau_m) a^{(m-1)}_{m}.\label{allCoefficients-k}
\end{align}
From those recursion relations, and initial conditions (\ref{initialaCondition})
we can find all the explicit expressions of $\coea{m}{k}$.
It is easy to see that
\begin{align}\label{allCoefficientsf-k}
  \coea{m}{1}=& \exp\left[-\frac{iE_1}{\hbar}\sum^m_{i=1}T_i \right]\cos(\Omega'_1\tau_1),  \\
  \coea{m}{2}=&-\exp\left[-\frac{iE_2}{\hbar}\sum^m_{i=1}T_i \right] 
  \cos(\Omega'_2\tau_2)i\sin(\Omega'_1\tau_1),
\end{align}
\begin{align}
  \coea{m}{m+1}=&\left[-e^{-iE_{m+1}T_m/\hbar}
  i\sin(\Omega'_{m}\tau_m)\right] \coea{m-1}{m}    \nonumber    \\
  = &  \cdots \nonumber \\
  = & \left(-i\right)^{m-1}\exp\left[-\frac{i}{\hbar}
      \sum^m_{i=2}E_{i+1}T_i\right]\prod^m_{i=2}\sin(\Omega'_i\tau_i) \coea{1}{2} \nonumber\\
  =&\exp\left[-\frac{i}{\hbar}\sum^m_{i=1}E_{i+1}T_i-
    i\frac{\pi}{2}m\right]\prod^m_{i=1}\sin(\Omega'_i\tau_i).
  \label{coefficient-m-m+1-2}
\end{align}
Then using \eqref{induction-model2-1} and $a^{(m-1)}_m$, which is obtained
from (\ref{coefficient-m-m+1-2}) by replacing $m$ by $m-1$, we have
\begin{align}\label{generalForm-second}
  \coea{m}{k}=& e^{-iE_k(\tau_m+\tau'_m)/\hbar}\coea{m-1}{k}  \nonumber \\
   = & e^{-iE_k(\tau_m+\tau'_m)/\hbar}e^{-iE_k(\tau_{m-1}+\tau'_{m-1})/\hbar} \coea{m-2}{k}   \nonumber \\
   = & \cdots   \nonumber \\
   = & \exp\left[-\frac{iE_k}{\hbar}\sum^m_{i=k+1}T_i\right] \coea{k}{k}  \nonumber \\
   = & \exp\left[-\frac{iE_k}{\hbar}\sum^m_{i=k+1}T_i\right]e^{-iE_k T_k \hbar}
     \cos(\Omega'_k\tau_k) \coea{k-1}{k}  \nonumber \\
   =& \exp\left[-\frac{iE_k}{\hbar}\sum^m_{i=k}T_i-\frac{i}{\hbar}\sum^{k-1}_{i=1}E_{i+1}T_i \right.   \nonumber \\
    & \left. -i\frac{\pi}{2}(k-1) \right]
      \cos(\Omega'_k\tau_k)\prod^{k-1}_{i=1}\sin(\Omega'_i\tau_i).
\end{align}
One can check that \eqref{generalForm-second} includes the case $k=2$ and $k=m$
as special cases.

Therefore, after $N-1$ cycles, we arrive at the target state
\begin{align}\label{finalFormOfcoefficient}
\coea{N-1}{1}=&\exp\left[-\frac{iE_1}{\hbar}\sum^{N-1}_{i=1}T_i\right]\cos(\Omega'_1\tau_1) \nonumber \\
\coea{N-1}{m}=& \exp\left[-\frac{iE_m}{\hbar}
        \sum^{N-1}_{i=m}T_i-\frac{i}{\hbar}\sum^{m-1}_{i=1}E_{i+1}T_i \right.\nonumber \\
  & \left.-i\frac{\pi}{2}(m-1)\right]\cos(\Omega'_m\tau_m)
    \prod^{m-1}_{i=1}\sin(\Omega'_i\tau_i), \nonumber \\
  & \quad (2\leq m\leq N-1), \nonumber \\
\coea{N-1}{N}=& \exp\left[-\frac{i}{\hbar}\sum^{N-1}_{i=1}E_{i+1}T_i-i\frac{\pi}{2}(N-1)\right]\nonumber \\
  & \times \prod^{N-1}_{i=1}\sin(\Omega'_i\tau_i).
\end{align}

\subsection{Control parameters}

To determine control parameters $\tau_i,\tau'_i, 1\leq i\leq N-1$, we write
the target state as
\begin{equation}\label{stateSuperposition}
     \ket{\psi}=\sum^N_{n=1}a_n \ket{n}=\sum^N_{n=1}\gamma_n C_n \ket{n},
\end{equation}
in which
\begin{align}\label{realProbabilityAmplitude}
     C_1&=\cos(\Omega'_1\tau_1)  \nonumber \\
     C_m&=\cos(\Omega'_m\tau_m)\prod^{m-1}_{i=1}\sin(\Omega'_i\tau_i),\quad (2\leq m\leq N-1), \nonumber\\
     C_N&=\prod^{N-1}_{i=1}\sin(\Omega'_i\tau_i),
\end{align}
and phase $\gamma_n$
\begin{align}\label{relativePhase}
   \gamma_1&= \exp\left[-\frac{iE_1}{\hbar}\sum^{N-1}_{i=1}T_i\right],  \nonumber \\
   \gamma_m&= \exp\left[-\frac{iE_m}{\hbar}
              \sum^{N-1}_{i=m}T_i-\frac{i}{\hbar}
              \sum^{m-1}_{i=1}E_{i+1}T_i\right.
             \left.-i\frac{\pi}{2}(m-1)\right], \nonumber  \\
   \gamma_N&= \exp\left[-\frac{i}{\hbar}
              \sum^{N-1}_{i=1}E_{i+1}T_i-i\frac{\pi}{2}(N-1)\right].
\end{align}

For a given target state, namely, $C_n$ and $\gamma_n$ are given, we can determine
the control parameters $\{\tau_n,\tau'_n|n=1,2,...,N-1  \}$. From $C_1$ we can determine
$\tau_1$, and then $\tau_2$ from $C_2$ and the obtained $\tau_1$. Recursively we can
obtain all $\tau_n$.

For $\tau^\prime_n$, from $\gamma_2$ and $\gamma_3$, we obtain
\begin{align}
\sum_{i=1}^{N-1} T_i, \qquad E_3 \sum_{i=2}^{N-1}+E_2 T_1,
\end{align}
respectively. As $E_2\neq E_3$, we obtain $T_1$ and $\sum_{i=2}^{N-1}T_i$.
From $\gamma_4$, we can obtain
\begin{equation}
E_4 \sum_{i=3}^{N-1}T_i + E_3 T_2
\end{equation}
from which we obtain $T_2$ as well as $\sum_{i=3}^{N-1}T_i$. Repeating this
process, we can obtain all $T_i$ and thus $\tau^\prime_i$.

\section{Conclusion}

In this paper we proposed a protocol to drive two types of  finite
dimensional quantum system to an arbitrary given target states. The control
parameters are time periods $\{\tau_m, \tau^\prime_m | m=1,2,\cdots,N-1\}$
which can be explicitly determined from the probability
amplitudes of the given target states. Relationship between control parameters
and amplitudes is triangular functions and can be solved explicitly. The
control fields in this protocol is the usual electric field described by
the cosine function.

We have $2(N-1)$ real control parameters. In the target state there are $N$
complex or $2N$ real parameters. Taking into account the normalization
condition of target state, one has $2(N-1)$ real parameters, the same as
the number of the real control parameters. From this fact we can
conclude that we can drive the system to an arbitrary target state by
choosing appropriate control parameters $\{\tau_m, \tau^\prime_m\}$.

As further works, we would like to consider the indirect control protocol of
finite quantum system by generalizing the control scheme in this paper. We
also would like to consider the control protocol in the presentence of
environment.

\section*{Ackonwledgement}

This work is supported by the National Science Foundation of China under grand
number 11075108.


\begin{thebibliography}{99}
\bibitem{huang} G. M. Huang, T. J. Tarn and J. W. Clark, {\it J. Math. Phys.} {\bf 24} 2608 (1983)

\bibitem{optimal} M. A. Daleh, A. M. Peirce, and H. Rabitz, {\it Phys. Rev. A} {\bf 37} 4950 (1988); \\
  A. Bartana, R. Kosloff, and D. J. Tannor, {\it Chem. Phys.} {\bf 267}, 95 (2001); \\
  U. Boscain, G. Charlot, J.-P. Gauthier, S. Duerin and H.-R. Jauslin,
  {\it J. Math. Phys.} {\bf 43} 2107 (2002).

\bibitem{controllability} V. Ramakrishna and H. Rabitz, {\it  Phys. Rev. A} {\bf 54} 1715 (1996).

\bibitem{fu1} S. G. Schirmer, H. Fu and A. I. Solomon, {\it Phys. Rev. A.} {\bf 63}, 063410 (2001).

\bibitem{fu2} H. Fu, S. G. Schirmer and A. I. Solomon, {\it Phys. A: Math. Gen.} {\bf 34} (2001).



\bibitem{graph} G. Turinici, Mathematical Models and Methods for ab Initio Quantum Chemistry
(Lecture Notes in Chemistry vol 74) ed. M. Defranceschi and C. Le Bris (Berlin: Springer, 2000); \\
G. Turinici and H. Rabitz, {\it Chem. Phys.} 267 (2001)

\bibitem{feedback} H. M. Wiseman and G. J. Milburn, {\it Quantum measurement and control}, Cambridge Press,
2010.

\bibitem{indirect1} H. C. Fu, H. Dong, X. F. Liu and C. P. Sun, {\it Phys. Rev. A.} {\bf 75} 052317 (2007).
\bibitem{indirect2} H. C. Fu, H. Dong, X. F. Liu and C. P. Sun, {\it J. Phys. A.} {\bf 42} 045303 (2009).

\bibitem{romano} R. Romano and D. D¡¯Alessandro, {\it Phys. Rev. Lett.} {\bf 97} 080402 (2006);
{\it Phys. Rev. A} {\bf 73} 022323 (2006).

\bibitem{penchen} A. Pechen and H. Rabitz, {\it Phys. Rev. A} {\bf 73} 062102 (2006).

\bibitem{cartan} D. D¡¯Alessandro1 and F. Albertini {\it J. Phys. A.} {\bf 40} 2439 (2007).




\bibitem{b} M. O. Scully, M. S. Zubairy,  {\it Quantum Optics}, 2000.
\end{thebibliography}
\end{document}